\documentclass[preprint2]{aastex}

\usepackage{lineno}

\shorttitle{The He CEW scenario for SN 2020eyj} \shortauthors{Meng \& Podsiadlowski}

\begin{document}


\title{The helium common-envelope wind scenario for SN 2020eyj}


\author{Xiang-Cun Meng$^{\rm 1}$  \& Philipp Podsiadlowski$^{\rm 2, 3}$}
\affil{$^{\rm 1}$
International Centre of Supernovae (ICESUN), Yunnan Key Laboratory of Supernova Research, Yunnan Observatories, Chinese
Academy of Sciences (CAS), Kunming 650216, China\\
 $^{2}$London Centre for Stellar Astrophysics, Vauxhall, London, UK\\
 $^{3}$University of Oxford, St
Edmund Hall, Oxford, OX1 4AR, UK} \email{xiangcunmeng@ynao.ac.cn}
\email{podsi@hotmail.com}





\begin{abstract}
SN 2020eyj is the first type Ia supernova (SN Ia) showing the
signature of a compact helium-rich circumstellar material (CSM). Such
a large CSM is difficult to explain in a single-degenerate scenario
where the donor star is a helium star. Here we show that, under
certain conditions, it is possible that the transfer of helium leads
to a common envelope (CE) engulfing the system, similar to the
common-envelope wind model proposed by Meng \& Podsiadlowski (2017).
If in such a helium common-envelope wind (HeCEW) model the initial white
dwarf (WD) mass is larger than 1.1 $M_{\rm \odot}$ and the helium star
more massive than 1.8 $M_{\rm \odot}$, the mass of a helium
CE can be larger than 0.3 $M_{\rm \odot}$ prior to supernova explosion. 
The CE mass heavily depends on the initial parameters of the
binary system. A dynamical CE ejection event could occur shortly before the supernova, and then our model may naturally explain the properties of SN
2020eyj, specifically the massive He-rich CSM, its dim peak
brightness, low ejecta velocity and low birth rate.

\end{abstract}





\section{INTRODUCTION}\label{sect:1}
Although type Ia supernovae (SNe Ia) are so important in cosmology
(e.g., \citealt{RIE98}; \citealt{PER99}), their exact origin is still
unclear, and many potential progenitor models have been proposed
(\citealt{WANGB12}; \citealt{MAOZ14};
\citealt{SOKER19}; \citealt{RUIZLAPUENTE19}; \citealt{LIUZW23}; \citealt{SOKER24}; \citealt{RUITER25}). 
While such models enjoy a degree of
empirical support, they are still simultaneously confronted with
substantial problems, both from theoretical and observational
perspectives (\citealt{HOWEL11}; \citealt{MAOZ14}; \citealt{JHA19}).
Recently, a unique SN Ia, SN 2020eyj, showing a delayed interaction
with a large amount of helium-rich circumstellar material (CSM) has
been reported, which provides an unusual opportunity to constrain the
progenitor system of a particular SN Ia (\citealt{KOOL22}). Using 
the optically thick wind (OTW) model (\citealt{HAC96}), \citet{KOOL22}
suggested that SN 2020eyj originated from a single degenerate (SD)
system consisting of a WD + helium star. However, \citet{SOKER22}
subsequently challenged some of the arguments in \citet{KOOL22} and
proposed that this supernova is from a version of the core-degenerate
(CD) channel, in which two major common-envelope (CE) phases occurred
and the supernova was the product of the merger of a WD and the core
of a helium red giant.

\citet{KOOL22} describe the main properties of SN 2020eyj in details; here
we list the ones that are most relevant for our study. 1) The most remarkable
feature is that the supernova shows clear signatures for a helium-rich
CSM with which the supernova ejecta interact. Before the interaction,
the light curve is similar to a normal SN Ia, but with a lower
ejecta velocity and a lower peak brightness. These properties may indicate
a SN Ia from an initially massive carbon-oxygen (CO) or hybrid
carbon-oxygen-neon (CONe) WD (\citealt{NOM03};
\citealt{MENGXC18}). 2) During the interaction phase,  the spectra
show many similarities to the SN Ia-CSM, PTF 11kx, but with
strong helium emission lines (\citealt{DILDAY12};
\citealt{SILVERMAN13}). The CSM interaction in SN 2020eyj is also
confirmed by the detection of a radio counterpart, which can be explained
either by a 0.3-1.0 $M_{\rm \odot}$ CSM in a wind model or
0.36 $M_{\rm \odot}$ CSM  in a shell model within $10^{\rm 17}$
cm from the supernova.

\citet{SOKER13} argued that the canonical SD model cannot account for
such a large amount of CSM around PTF 11kx.  Similarly, the canonical
SD system of a WD + a helium star cannot explain the amount of CSM in
the case of SN 2020eyj, especially for the wind case considered in
\citet{KOOL22} (see also \citealt{SOKER22}).  For the wind model in
\citet{KOOL22}, the mass-transfer rate before the supernova explosion
would have to be at least $10^{\rm -3}$ $M_{\rm \odot}/{\rm yr}$, and
possibly as high as $3\times10^{\rm -2}$ $M_{\rm \odot}/{\rm yr}$,
which seems impossible for binary evolution models (\citealt{WANGB09};
\citealt{MORIYA19}). The reason for such a high inferred
mass-transfer/mass-loss rate is that in the OTW model
the wind velocity is higher than 1000 km/s (\citealt{HAC96};
\citealt{MORIYA19}).  However, many predictions from the OTW model are
not confirmed by other observations (see \citealt{MENGXC17} for
detailed discussions).  \citet{KOOL22}, also realizing these problems,
therefore suggested that there could be a long-lived disk around the
progenitor system, which may however raise other problems (see the discussion
in \citealt{SOKER22}).

\citet{MENGXC17} constructed a new version of the SD model, the common
envelope wind (CEW) model, which can explain the large amount of CSM
around SNe Ia-CSM (see also \citealt{MENGXC18} and \citealt{SOKER19b})
and avoids the mass-transfer rate problem in \citet{KOOL22}. In
particular, the wind velocity is lower than 100 km/s in the CEW model,
similar to that in the CD model (\citealt{CUIYZ02}). In this paper,
based on detailed binary evolution calculations of WD + He star
systems, we will show that in such systems a helium CE can form
just as in the CEW model and has the potential to explain the
properties of SN 2020eyj, if there still is a massive CE around the
progenitor system at the time of the explosion (which we will refer to
as a HeCEW model).

The paper is organized as following. In section \ref{sect:2}, we
describe our method and present the results of the calculations in
section \ref{sect:3}. We discuss the results and present our main
conclusions in section \ref{sect:4}.

\section{METHOD}
\label{sect:2}

The basic description of the HeCEW model here is similar to
the CEW model in \citet{MENGXC17}, except that the companion is a
helium star. We do not
repeat the details here and just describe the differeces. We
calculate the binary evolution of WD + He star systems, where the
mass transfer from the companions to the WDs may begin when the
companions are on the helium main sequence (MS) or in the  helium
Hertzsprung gap (HG).  Here, orbital angular momentum loss by
gravitational wave radiation (GWR) is included by adopting the
standard formula from \citet{LANDAU71}.

For a binary system consisting of a WD + a He star, after the He
star fills its Roche lobe,  a phase of Roche lobe overflow (RLOF) occurs
and the helium star transfers helium-rich material onto the
surface of the WD, which then increases its mass. If the WD
mass grows to 1.378 $M_{\rm \odot}$, we assume that the WD 
explodes as a SN Ia (\citealt{NTY84}).

The mass-growth rate of the WD is calculated as follows. 1) If
the mass-transfer rate from the helium companion to the WD
exceeds a critical accretion rate, $\dot{M}_{\rm cr}$,
we assume a CE forms around the binary system and that the WD
increases its mass according to $\dot{M}_{\rm WD}=\dot{M}_{\rm cr}$, where
the method to calculate the $M_{\rm CE}$ is the same to that in
\citet{MENGXC17}. The critical accretion rate is given 
by \citet{NOMOTO82} as

\begin{equation}
  \dot{M}_{\rm cr}=7.2\times10^{\rm -6}(M_{\rm WD}/M_{\rm \odot}-0.6) ~{M_{\rm \odot}/{\rm yr}}. \label{eq:1}
\end{equation}
2) If the CE disappears, the mass growth rate of the WD is given by
\begin{equation}
    \dot{M}_{\rm WD}=\eta_{\rm He}|\dot{M}_{\rm 2}|, \label{eq:2}
\end{equation}
where $\eta_{\rm He}$ is taken from \citet{KH04}. Then, the WD may explode
with or without a CE. If the CE mass is more massive than 0.3 $M_{\rm
  \odot}$ when $M_{\rm WD}=1.378$ $M_{\rm \odot}$, we assume that this
explosion is a candidate system for SN 2020eyj. 
Actually, the observational signature of the interaction between the supernova ejecta and the helium-rich CSM does not clearly emerge 
until approximately 50 days after the explosion. This necessarily requires that a CE ejection took place before the supernova explosion to 
form a gap between the progenitor system and the dense CSM.  We will discuss this in section \ref{sect:4}.

A relatively massive initial WD possesses a lower carbon abundance.  
This results in a lower production of $^{\rm 56}$Ni during the SN Ia explosion, 
leading to a lower peak luminosity (\citealt{NOM03}; \citealt{MENGYANG11}). 
Therefore, the relatively dim peak brightness of SN 2020eyj may imply that 
it originated from a relatively massive initial WD.  Since a hybrid
CONe WD could produce a SN Ia-CSM (\citealt{MENGXC18}), we assume here
that the initial WDs leading to SNe Ia could be as massive as 1.3
$M_{\rm \odot}$ and therefore set the initial WD mass to 1.1, 1.2 and
1.3 $M_{\rm \odot}$, respectively.  The initial masses of the donor
stars, $M^{\rm i}_{\rm 2}$, range from 1.5 to 3.4 $M_{\rm \odot}$ in
steps of 0.1 $M_{\rm \odot}$; the initial orbital periods, $P^{\rm
  i}$, are taken from the range of $\log (P^{\rm i}/{\rm d})=-1.2$ to
1.3 in steps of 0.1.  Our parameter space is smaller than in
\citet{WANGB09}, as we restrict it to the range where the WD can
potentially explode inside a CE.




\begin{figure*}
\centerline{\includegraphics[angle=270,scale=.65]{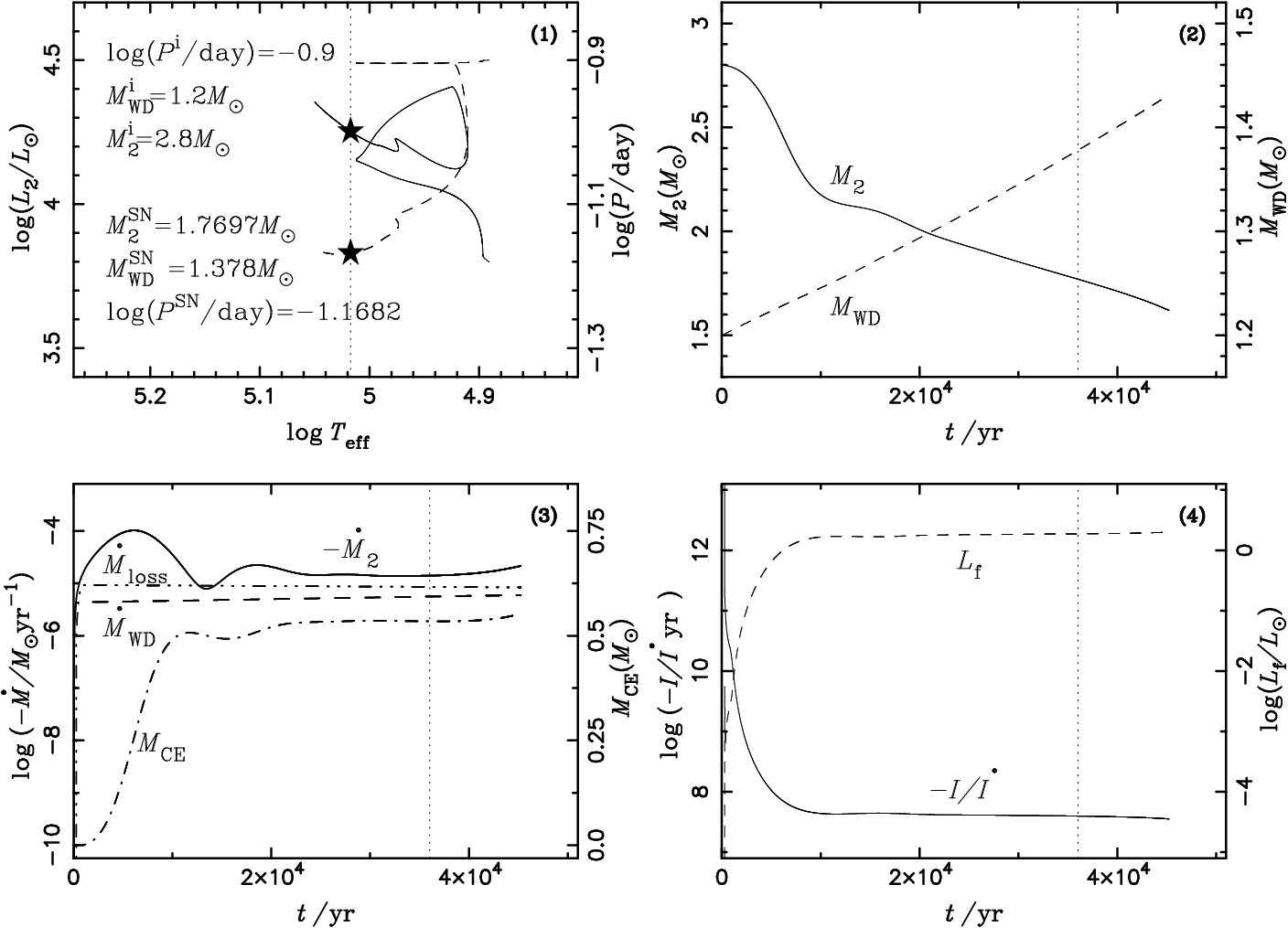}}
\caption{Illustrative binary evolution calculation in the HeCEW model. The evolution
    of various parameters is shown, including the WD mass,
    $M_{\rm WD}$, the secondary mass, $M_{\rm 2}$, the mass-transfer
    rate, $\dot{M}_{\rm 2}$, the mass-growth rate of the WD,
    $\dot{M}_{\rm WD}$, the mass of the CE, $M_{\rm CE}$, the
    mass-loss rate from the system, $\dot{M}_{\rm loss}$, the
    frictional luminosity, $L_{\rm f}$, and the merger timescale for
    the binary system, $-I/\dot{I}$, as labeled in each panel. The
    evolutionary track of the donor star and the evolution of the
    orbital period are shown as solid and dashed curves in panel (1),
    respectively. Dotted vertical lines in all panels and asterisks in
    panel (1) indicate the position where the WD is expected to
    explode as a SN Ia. The initial and the final binary parameters
    are given in panel (1).}\label{122809}
\end{figure*}

\begin{figure}
    \centerline{\includegraphics[angle=270,scale=.36]{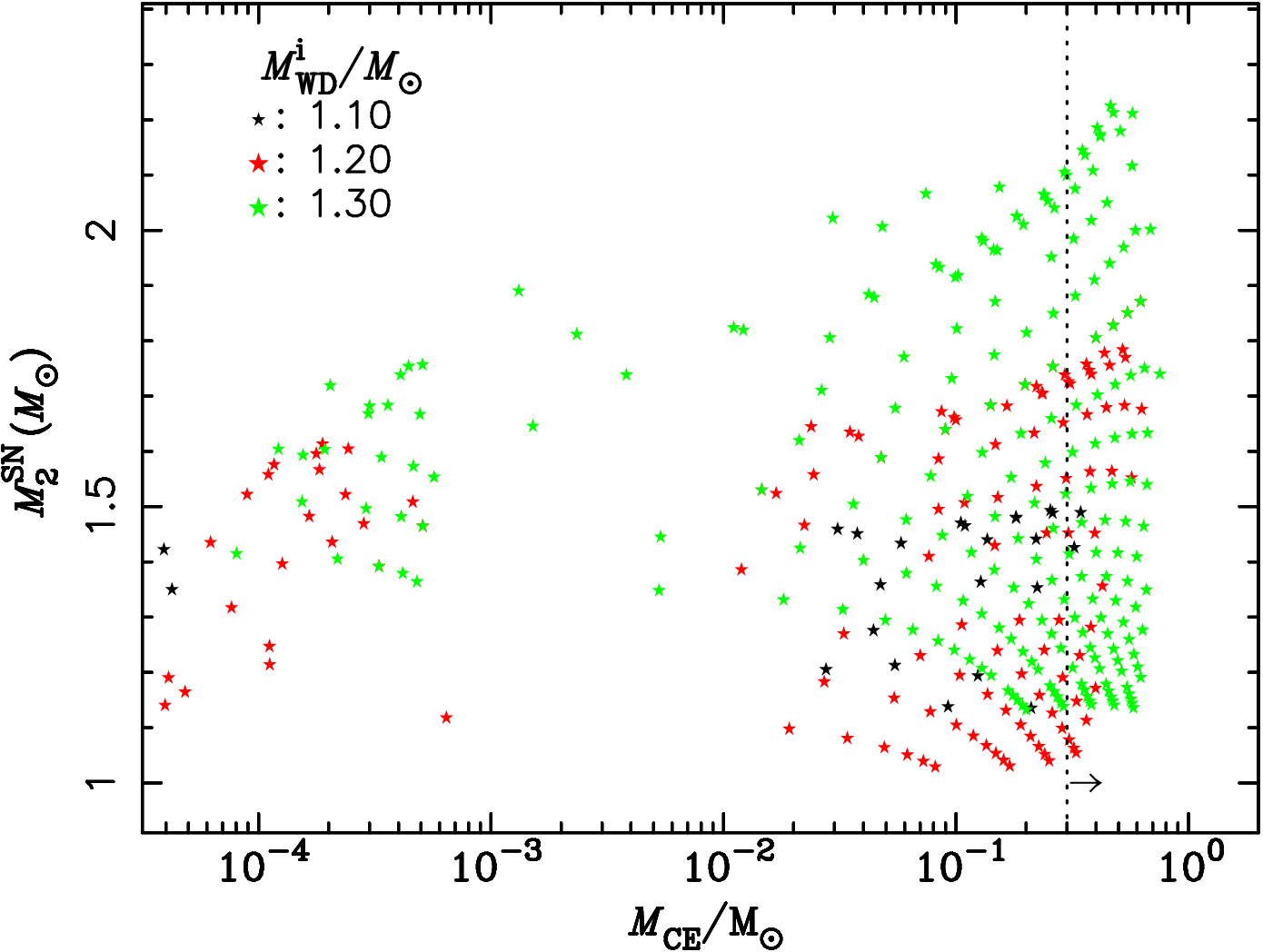}}
\caption{The CE mass and the companion mass when $M_{\rm
WD}=1.378$ $M_{\rm \odot}$ for different initial WD masses. The
vertical dotted line shows the lower limit for the amount of CSM
around the progenitor of SN 2020eyj.}\label{m2mce}
\end{figure}

\begin{figure}
    \centerline{\includegraphics[angle=270,scale=.36]{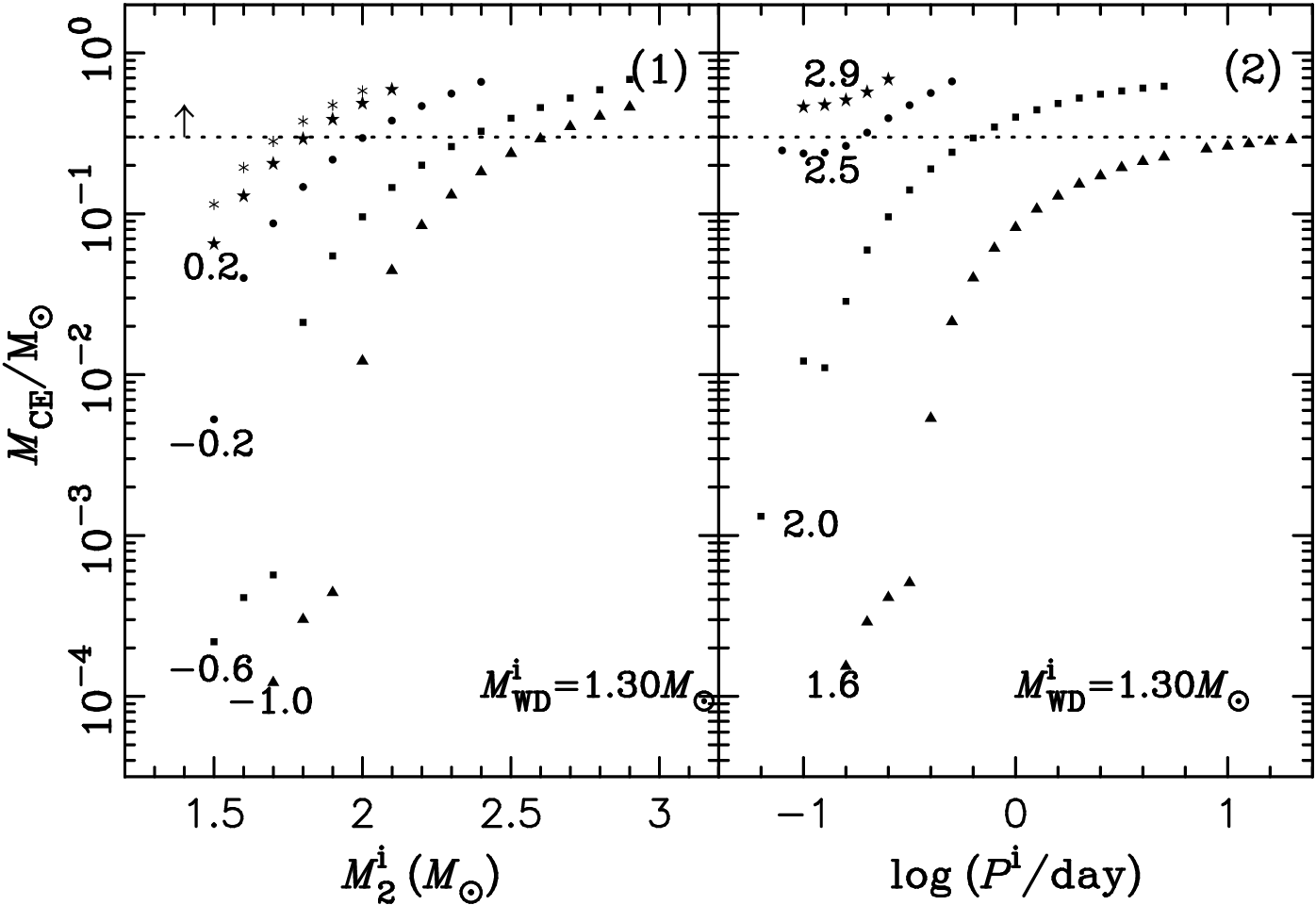}}
 \caption{The CE mass as a function of the initial companion mass
when $M_{\rm WD}=1.378$ $M_{\rm \odot}$ for $M^{\rm i}_{\rm
WD}=1.3$ $M_{\rm \odot}$. The  horizontal dotted line shows the
lower limit for the amount of CSM required around the progenitor of SN
2020eyj.}\label{mcepara}
\end{figure}

\section{RESULTS}\label{sect:3}
Fig.~\ref{122809} shows the evolution of the key binary parameters
from a WD + He star system, as well as the evolutionary track of the
donor star in the Hertzsprung-Russell (HR) diagram and the evolution
of the orbital period. In this example, the initial system has a donor
of $M^{\rm i}_{\rm 2}=2.8$ $M_{\rm \odot}$, a WD of $M^{\rm i}_{\rm
  WD}=1.2$ $M_{\rm \odot}$, and an initial orbital period of $\log
(P^{\rm i}/{\rm d})=-0.9$.  Actually, the binary evolution here is
similar to that shown in \citet{WANGB09}, although the basic physical
picture is different (see also \citealt{MENGXC17}). The companion
fills its Roche lobe in the helium HG, i.e.  the system experiences
case BB RLOF. The mass-transfer rate exceeds $\dot{M}_{\rm cr}$ soon
after the onset of RLOF and then a CE forms around the binary system,
where a part of the CE material is lost from the surface of the
CE. Because the mass-transfer rate is generally higher than
$\dot{M}_{\rm WD}+\dot{M}_{\rm loss}$, the CE mass increases to a
value exceeding 0.5 $M_{\rm \odot}$. After $3.6\times10^{\rm 4}$ yrs,
the WD increases its mass to 1.378 $M_{\rm \odot}$, where it
explodes. At this stage, the mass-transfer rate is still higher than
$\dot{M}_{\rm cr}$, and the CE has a mass of 0.5344 $M_{\rm
  \odot}$. Such a CE is massive enough to explain the large CSM around
SN 2020eyj.

Fig.~\ref{m2mce} shows the final outcomes of all the binary
evolution calculations for different initial WD masses in the $M_{\rm
CE}$ - $M^{\rm SN}_{\rm 2}$ plane. The figure clearly shows that,
at the moment when  $M_{\rm WD}=1.378$ $M_{\rm \odot}$, many models
have a CE with a mass larger than 0.3 $M_{\rm \odot}$, where the largest
mass is 0.78 $M_{\rm \odot}$. The figure also
shows that, the more massive the initial WD, the larger the
parameter space fulfilling the CSM constraint for SN 2020eyj.
Probably any system with $M_{\rm CE}$ larger than
$\sim0.1$ $M_{\rm \odot}$ when $M_{\rm WD}=1.378$ $M_{\rm \odot}$ would
look like a SN Ia-CSM. In
addition, similar to Fig. 18 in \citet{MENGXC17}, there seems a
gap between $M_{\rm CE}=10^{\rm -3}$ $M_{\rm \odot}$ and $10^{\rm
-2}$ $M_{\rm \odot}$, which originates from different
evolutionary stages of the companion stars at the onset of RLOF
which sets the mass-transfer time-scale (see details in
\citealt{MENGXC17}).

Fig.~\ref{mcepara} shows how the CE mass at the time of the supernova
depends on the initial binary parameters for the case of $M^{\rm i}_{\rm WD}=1.3$ $M_{\rm \odot}$, 
for which the initial parameter
space for SN 2020eyj is the largest. The figure shows that, for
fixed initial WD mass and fixed initial orbital
period, the CE mass at the time of the supernova increases with the
initial companion mass. For a companion mass 
more massive than 1.8 $M_{\rm \odot}$, the explosion can produce the
required properties to explain SN 2020eyj. Similarly, for the system
with a given initial WD mass and a given companion mass, the CE mass
at the time of the supernova generally increases with the initial
orbital period. To produce a SN 2020eyj-like supernova, the
initial period should be longer than $\log (P^{\rm i}/{\rm d})=-1.0$, 
and could be as long as $\log (P^{\rm i}/{\rm d})=1.3$, for the
case of $M^{\rm i}_{\rm WD}=1.3$ $M_{\rm \odot}$.  



For our HeCEW model, a large amount of helium-rich material may be
lost from the CE surface to form the CSM, and the total amount could
be as massive as $\sim 1$ $M_{\rm \odot}$, lower than that from the WD
+ MS channel by a factor of $\sim 3$ (\citealt{MENGXC17}). Except for
the helium lines in the spectra, such CSM may also form dust and cause
light echoes (\citealt{WANGXF08}; \citealt{YANGY17}). We can simply
calculate the maximum distance that the lost material will have
reached when $M_{\rm WD}=1.378$ $M_{\rm \odot}$ by $d=v_{\rm W}\times
t_{\rm d}$, where $v_{\rm W}$ is the wind velocity of the material
lost, and $t_{\rm d}$ is the delay time from the onset of mass
transfer to the time of the supernova.  
Assuming a wind velocity of 50 km/s, a plausible value for the wind from a He CE, 
the maximum distance derived here is shorter than that for WD + MS systems 
reported by \citet{MENGXC17} by a factor of about 80. This difference stems primarily 
from the difference in the mass-growth rates of the WD between the WD + He star and WD + MS channels. 
Specifically, the mass-growth rate during the CE phase in the WD + He star channel is roughly 7 times higher 
than that in the WD + MS channel. An additional reason is that, in this study, 
we only consider the systems with relatively massive companions, in which all WDs explode during the CE phase. 
In contrast, \citet{MENGXC17} focused on systems that may contain less massive MS companions, 
allowing the WDs to explode in the so-called recurrent nova phase, corresponding to a mass-transfer timescale on the order of a few $10^6$ years.
These two reasons lead to a significantly shorter delay time ($t_{\rm d}$) for the WD + He star systems examined in this work compared to the WD + MS ones in \citet{MENGXC17}.

\begin{figure}
    \centerline{\includegraphics[angle=270,scale=.36]{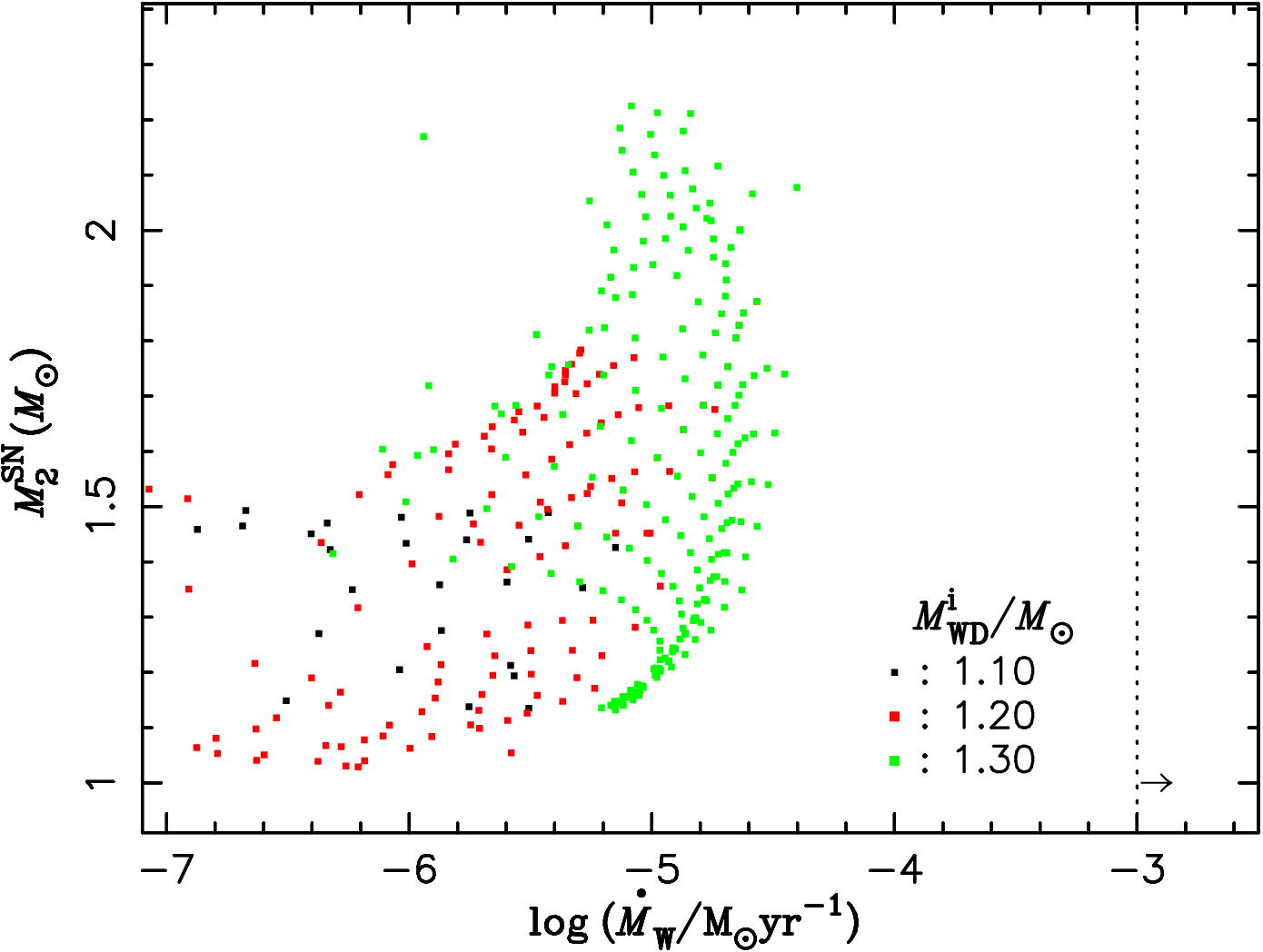}}

 \caption{The wind mass-loss rate and the companion mass when
$M_{\rm WD}=1.378$ $M_{\rm \odot}$ for different initial WD
masses. The vertical dotted line shows the lower limit of the wind
mass-loss rate for the progenitor of SN 2020eyj if the OTW
model were adopted.}\label{m2mdot}
\end{figure}



\section{DISCUSSIONS AND CONCLUSIONS}\label{sect:4}
This work was motivated by the discovery of SN 2020eyj, a SN Ia-CSM
with a large helium-rich CSM (\citealt{KOOL22}). Although
\citet{KOOL22} favour the SD scenario, some of the detailed physical
aspects may be problematic. For example, based on the OTW model, they
adopted, the mass-loss rate needed would have to be higher than
$10^{\rm -3}$ $M_{\rm \odot}/{\rm yr}$, possibly as high as
$3\times10^{\rm -2}$ $M_{\rm \odot}/{\rm yr}$.  Fig.~\ref{m2mdot},
shows estimates for the wind mass-loss rate and the companion mass
when $M_{\rm WD}=1.378$ $M_{\rm \odot}$ for different initial WD
masses for the OTW model, where $\dot{M}_{\rm W}=|\dot{M}_{\rm
  2}|-\dot{M}_{\rm WD}$ \footnote{The binary evolution in the OTW and
CEW models is similar, although some of the physical assumptions are
different. Here we use the $|\dot{M}_{\rm 2}|-\dot{M}_{\rm WD}$ from
our HeCEW model to estimate the wind mass-loss rate from the OTW
model. Such a wind mass-loss rate is slightly larger than that from
the OTW due to the slightly higher $|\dot{M}_{\rm 2}|$.}.  The figure
clearly shows that, even for the most optimistic case, the wind
mass-loss rate from the OTW WD + He star scenario is lower than that
needed by SN 2020eyj by a factor of $\sim25-750$.  If the mass-loss
rate is really as high as suggested by \citet{KOOL22}, the system
would be in a CE phase, as suggested in this paper and in
\citet{SOKER22}.

In this paper, we found that, only when the initial WD is massive
enough, the progenitor system consisting of a WD + a He star may
produce a massive He-rich CSM, i.e. $M^{\rm i}_{\rm WD}\geq 1.1$ $M_{\rm
\odot}$. Considering the correlation between the initial WD mass
and the peak brightness of SNe Ia (\citealt{NOM03}; \citealt{MENGYANG11}), our model may
naturally explain the dim peak brightness of SN 2020eyj, as well
as the low ejecta velocity. In other words, our model predicts
that the SN 2020eyj-like supernovae have a relatively dim peak
brightness, if the interaction between the supernova ejecta and
the CSM occurs after the maximum light.

We did not carry out a detailed binary population synthesis
simulation to estimate the rate of SN 2020eyj-like supernovae.
However, we can get a rough estimate of the birth rate based on previous
detailed calculations.  The WD + He star channel may contribute to
about 1\% - 10\% of all SNe Ia (\citealt{WANGB09b,
  WANGB17}). Considering that about 1 in 100 SNe Ia explodes in a
massive CE, this yields an estimate of about 0.01\% - 0.1\% of SNe Ia
belonging to the class of SN 2020eyj-like supernovae. This relative
rareness may explain why only one such supernova has been observed
until now.

Observations of SN 2020eyj reveal a time delay of tens of days between the supernova explosion and its subsequent interaction with the CSM, 
indicating the presence of a gap between its progenitor and the dense CSM. This implies the need for a Merger-to-Explosion Delay (MED) in our model (see \citealt{SOKER22b,SOKER24b} for details). \citet{SOKER22} estimate a MED of $\sim 10$ yr for SN 2020eyj.  Although the physical mechanisms responsible for forming such a gap or for the MED remain unclear (\citealt{MENGXC13};
\citealt{SOKER22b}), a magnetized and rapidly rotating WD may play a role in mediating the delay (\citealt{JUSTHAM11}; \citealt{ILKOV12}; \citealt{NEOPANE22}). In our canonical CEW model, the CE is always dynamically unstable, 
suggesting that a CE ejection event could occur at any time prior to a supernova explosion (\citealt{CUIYZ02}). 
Similarly, a helium-rich CE is also likely to be dynamically unstable, as in the canonical CEW scenario. 
A dynamical ejection event occurring shortly before the supernova could thus create the gap required by the observations of SN 2020eyj. Although a detailed dynamical simulation is necessary in the future to investigate the exact MED from the HeCEW scenario, a MED of a few years could be possible (\citealt{CUIYZ02}).

In summary, we demonstrates the feasibility of the HeCEW model 
in explaining helium-rich SNe Ia-CSM such as SN 2020eyj.
A more comprehensive and detailed investigation of the HeCEW model will be presented in future work.

\section*{Acknowledgments}
We thank Dongdong Liu for his kind help and discussion. This work is supported by the National Natural Science Foundation of China (Nos. 12288102, 12333008 and 11973080),  the National Science Foundation of China and National Key R\&D Program of China (No. 2021YFA1600403) and the Strategic Priority Research
Program of the Chinese Academy of Sciences (grant Nos. XDB1160303, XDB1160000). X.M. acknowledges support from International Centre of Supernovae, Yunnan Key Laboratory (No. 202302AN360001), Yunnan Fundamental Research Projects (NOs. 202401BC070007 and 202201BC070003), the Yunnan Revitalization Talent Support Program-Science \& Technology Champion Project (NO. 202305AB350003),  Chinese Academy of Sciences President's  International Fellowship Initiative Grant No. 2026PVA0010 and the science research grants from the China Manned Space Project with grant no. CMS - CSST - 2025 - A13.

\end{document}